\documentclass[manuscript]{acmart}
\usepackage{hyperref}
\usepackage[version=4]{mhchem}
\usepackage{xcolor}
\usepackage{cleveref}

\copyrightyear{2026}
\acmYear{2026}
\setcopyright{cc}
\setcctype{by}
\acmConference[HPDC '26]{The 35th International Symposium on High-Performance Parallel and Distributed Computing}{July 13--16, 2026}{Cleveland, OH, USA}
\acmBooktitle{The 35th International Symposium on High-Performance Parallel and Distributed Computing (HPDC '26), July 13--16, 2026, Cleveland, OH, USA}
\acmDOI{10.1145/3806645.3816157}
\acmISBN{979-8-4007-2640-8/2026/07}

\begin{document}

%%
%% The "title" command has an optional parameter,
%% allowing the author to define a "short title" to be used in page headers.
\title{Expressibility, Noise, and Error Mitigation in VQE Ansatz Selection}

%%
%% The "author" command and its associated commands are used to define
%% the authors and their affiliations.
%% Of note is the shared affiliation of the first two authors, and the
%% "authornote" and "authornotemark" commands
%% used to denote shared contribution to the research.
\author{Peter Annis}
\email{pannis@mail.umw.edu}
\author{Abe Kassem}
\email{akassem2@mail.umw.edu}
\author{Evan Coleman}
\email{ecolema4@umw.edu}
\affiliation{%
  \institution{University of Mary Washington}
  \city{Fredericksburg}
  \state{Virginia}
  \country{USA}
}

\renewcommand{\shortauthors}{Annis, Kassem, Coleman}

\begin{abstract}
The variational quantum eigensolver (VQE) is a promising algorithm for near-term quantum chemistry applications, but selecting optimal ansatz circuits remains challenging. Expressibility, a metric quantifying a circuit's ability to explore the Hilbert space, has been proposed as a guide for ansatz selection, but recent work showed it inconsistently predicts VQE performance under realistic noise for \ce{H2}. We extend this investigation to cover both \ce{H2} and \ce{H3+} under four execution scenarios: ideal, noisy, and noisy with zero-noise extrapolation (ZNE) or probabilistic error cancellation (PEC). We find that error mitigation does not reliably restore expressibility's predictive power. ZNE reduces error for only 4 of 12 \ce{H2} circuits and 4 of 6 \ce{H3+} circuits, while PEC actually increases error in 11 of 12 \ce{H2} circuits and all 6 \ce{H3+} circuits. We reproduce and extend Saib et al.'s key finding that circuit rankings scramble under noise (Spearman $\rho \approx -0.1$ between ideal and noisy rankings), and identify a new result: ZNE largely preserves noisy rankings ($\rho = +0.80$ for \ce{H2}) while PEC actively reorders them ($\rho = -0.22$). Noisy expressibility, computed from density matrix simulations, strongly predicts unmitigated performance for \ce{H3+} (Pearson $r = +0.91$, $p = 0.01$), but this metric is computationally intractable at scale. We demonstrate that zero-cost circuit topology metrics such as two-qubit gate count provide comparable or superior predictive power for PEC degradation ($r = +0.96$ for \ce{H3+}), while standard expressibility best predicts noisy and ZNE performance for \ce{H2} ($r = +0.74$ and $r = +0.77$).
\end{abstract}

%%
%% The code below is generated by the tool at http://dl.acm.org/ccs.cfm.
%%


%%
%% Keywords. The author(s) should pick words that accurately describe
%% the work being presented. Separate the keywords with commas.

% \received{TBD 2026}
% \received[revised]{TBD 2026}
% \received[accepted]{TBD 2026}

\maketitle

%==============================================================================
\section{Introduction}
%==============================================================================

The variational quantum eigensolver (VQE) has emerged as a leading algorithm for quantum chemistry simulations on noisy intermediate scale quantum (NISQ) devices \cite{peruzzo2014variational,mcclean2016theory,kandala2017hardware}. By combining parameterized quantum circuits with classical optimization, VQE can approximate molecular ground state energies with shallow circuits suitable for near-term hardware. However, the performance of VQE critically depends on the choice of ans{a}tz.
Selecting an appropriate ans{a}tz presents a fundamental challenge: hardware-efficient ans{\"a}tze \cite{kandala2017hardware} minimize circuit depth but may lack the expressibility needed to represent target states accurately, while problem-inspired ans{\"a}tze like unitary coupled cluster (UCC) \cite{romero2018strategies} offer better chemical accuracy but require deeper circuits that accumulate more noise. This trade-off necessitates quantitative metrics to guide ans{a}tz selection for specific molecules and hardware platforms.

Expressibility has been proposed as such a metric, quantifying how uniformly a parameterized circuit can sample states across the Hilbert space \cite{sim2019expressibility}. Higher expressibility theoretically enables better approximation of target states. However, Saib et al. \cite{saib2021effect} demonstrated that expressibility exhibits weak correlation with VQE performance under noise (Pearson $r = 0.012$ for \ce{H2}), questioning its utility as a practical selection criterion. This finding was limited to a single molecule (\ce{H2}) and did not consider error mitigation techniques. Modern VQE implementations routinely employ quantum error mitigation (QEM) techniques such as zero-noise extrapolation (ZNE) \cite{temme2017error,giurgica2020digital} and probabilistic error cancellation (PEC) \cite{temme2017error,van2023probabilistic} to improve accuracy. These methods aim to recover ideal circuit behavior from noisy measurements. If successful, error mitigation could restore conditions under which ideal expressibility becomes predictive of performance. To date, no study has investigated whether QEM techniques rehabilitate expressibility as a useful ans{a}tz selection metric, nor whether the interaction between mitigation effectiveness and circuit properties reveals new predictors of VQE success.

In this work, we systematically investigate the relationship between expressibility and VQE performance for \ce{H2} (12 circuits, 4 qubits) and \ce{H3+} (6 circuits, 6 qubits) under four execution scenarios: ideal, noisy without mitigation, noisy with ZNE, and noisy with PEC. We test the hypothesis that error mitigation restores the correlation between ideal expressibility and VQE accuracy. Contrary to this hypothesis, we find that mitigation frequently degrades performance and that the predictive utility of circuit metrics is condition-dependent. We also compare expressibility against zero-cost circuit topology metrics, finding that no single predictor dominates across all conditions, and that expressibility loses discriminative power at 12+ qubits as circuits approach 2-designs. We also show preliminary scaling results for \ce{LiH} (4 circuits, 12 qubits), and \ce{BeH2} (2 circuits, 14 qubits).
The remainder of this paper is organized as follows. Section~\ref{sec:related} surveys related work on expressibility, noise effects, and error mitigation. Section~\ref{sec:background} provides technical background on VQE, expressibility, and QEM. Section~\ref{sec:methods} describes our experimental methodology. Section~\ref{sec:results} presents our results on ranking instability, mitigation effectiveness, and the comparative predictive power of circuit metrics. Section~\ref{sec:discussion} interprets these findings, and Section~\ref{sec:conclusion} concludes with recommendations for practitioners and future work.

%==============================================================================
\section{Related Work}
\label{sec:related}
%==============================================================================

Sim, Johnson, and Aspuru-Guzik~\cite{sim2019expressibility} introduced expressibility as a quantitative measure of a parameterized quantum circuit's ability to uniformly sample states across Hilbert space. %The metric compares the distribution of fidelities between randomly sampled circuit output states to the Haar random distribution using Kullback-Leibler divergence, where lower values indicate more uniform coverage. The authors also introduced entangling capability as a complementary metric based on the Meyer-Wallach measure. Crucially, they noted that ``the correlation between each descriptor and performance of an algorithm remains to be investigated''; a gap our work directly addresses.
Several alternative characterizations have since emerged. Du et al.~\cite{du2022efficient} proposed using the covering number as an expressivity measure, observing that expressivity decays exponentially with circuit depth under noise. Haug, Bharti, and Kim~\cite{haug2021capacity} introduced effective quantum dimension using quantum Fisher information geometry, identifying a transition region where circuits are expressive yet avoid barren plateaus. Most recently, Brozzi et al.~\cite{brozzi2025hamiltonian} introduced Hamiltonian expressibility (a problem-specific variant measuring a circuit's ability to explore a particular Hamiltonian's energy landscape), finding this metric better predicts performance for specific problem classes, particularly under noisy conditions.

A fundamental obstacle to leveraging highly expressive circuits emerged with McClean et al.'s~\cite{mcclean2018barren} discovery of barren plateaus in quantum neural network training landscapes. They proved that for parameterized circuits forming approximate 2-designs, gradient variance decays exponentially as $O\left(2^{-2n}\right)$ with qubit count $n$, rendering random parameter initialization unsuitable for optimization.
Holmes et al.~\cite{holmes2022connecting} made the connection to expressibility explicit, deriving a fundamental relationship between expressibility and trainability. Their key result establishes that highly expressive ans\"{a}tze exhibit flatter cost landscapes and are harder to train. Cerezo et al.~\cite{cerezo2021cost} further showed that barren plateaus arise even in shallow circuits when using global cost functions. Grimsley et al.~\cite{grimsley2023adaptive} showed ADAPT-VQE avoids barren plateaus by design through gradient-informed, one-operator-at-a-time construction. Tang et al.~\cite{tang2021qubit} demonstrated qubit-ADAPT-VQE reduces circuit depth by an order of magnitude compared to fermionic-ADAPT.

The comprehensive review by Larocca et al.~\cite{larocca2025review} synthesizes these findings, providing a unified taxonomy of barren plateau sources including ansatz design, initial states, observables, loss functions, and hardware noise. 
The hardware-efficient VQE demonstration by Kandala et al.~\cite{kandala2017hardware} established the paradigm of tailoring ans{\"a}tze to available hardware interactions to minimize circuit depth and noise accumulation. 

Systematic studies have since characterized how different noise types affect VQE. Zeng et al.~\cite{zeng2021simulating} found that ground state energy deviates from exact values as noise probability increases. Fontana et al.~\cite{fontana2022evaluating} demonstrated a linear relationship between mean energy values and noise intensity, and Dalton et al.~\cite{dalton2024quantifying} established critical quantitative thresholds for achieving a desired accuracy.
Wang et al.~\cite{wang2021noise} proved that local Pauli noise causes training landscapes to exhibit barren plateaus regardless of circuit structure, and Singkanipa and Lidar~\cite{singkanipa2025beyond} extended these results to non-unital noise.

The central finding motivating our work comes from Saib, Wallden, and Akhalwaya~\cite{saib2021effect}, who investigated the correlation between expressibility and VQE performance for \ce{H2} under noise. Their simulations revealed weak correlation (Pearson $r = 0.012$) between expressibility and energy difference under noisy conditions, compared to $r = -0.195$ under ideal conditions. They concluded that expressibility is not an effective measure of performance. Notably, the ranking of optimal ans{\"a}tze changed significantly in the presence of noise, and different IBM quantum devices within the same hardware family required different optimal ans{\"a}tze.

%==============================================================================
\section{Background}
\label{sec:background}
%==============================================================================

%------------------------------------------------------------------------------
\subsection{Variational Quantum Eigensolver}
\label{subsec:vqe}
%------------------------------------------------------------------------------

The variational quantum eigensolver (VQE) is a hybrid quantum-classical algorithm for approximating the ground state energy of a quantum system~\cite{peruzzo2014variational,mcclean2016theory}. Given a Hamiltonian $H$ describing a molecular system, VQE exploits the variational principle: for any normalized trial state $|\psi\rangle$, the expectation value of $H$ provides an upper bound on the true ground state energy $E_0$,$
    E(\boldsymbol{\theta}) = \langle\psi(\boldsymbol{\theta})|H|\psi(\boldsymbol{\theta})\rangle \geq E_0.
$

The trial state $|\psi(\boldsymbol{\theta})\rangle = U(\boldsymbol{\theta})|0\rangle^{\otimes n}$ is prepared by applying a parameterized quantum circuit (the ansatz) $U(\boldsymbol{\theta})$ to an initial reference state, typically $|0\rangle^{\otimes n}$. The parameter vector $\boldsymbol{\theta} \in \mathbb{R}^p$ contains the rotation angles for parameterized gates.
The molecular Hamiltonian is expressed as a linear combination of Pauli operators via fermion-to-qubit mappings such as Jordan-Wigner or Bravyi-Kitaev, $H = \sum_{\alpha} h_\alpha P_\alpha$,
where $h_\alpha \in \mathbb{R}$ are coefficients determined by the molecular geometry and basis set, and $P_\alpha \in \{I, X, Y, Z\}^{\otimes n}$ are Pauli strings. The energy expectation value is then $
    E(\boldsymbol{\theta}) = \sum_{\alpha} h_\alpha \langle\psi(\boldsymbol{\theta})|P_\alpha|\psi(\boldsymbol{\theta})\rangle$,
where each term $\langle P_\alpha \rangle$ is estimated by repeated measurement on the quantum processor.

The VQE algorithm iterates between quantum state preparation and measurement, and classical parameter optimization. A classical optimizer updates $\boldsymbol{\theta}$ to minimize $E(\boldsymbol{\theta})$, with common choices including gradient-based methods (SPSA, Adam) and gradient-free methods (COBYLA, Nelder-Mead). Convergence yields parameters $\boldsymbol{\theta}^*$ such that $E(\boldsymbol{\theta}^*)$ approximates $E_0$.

The choice of ansatz critically affects VQE performance. For example, hardware-efficient ans{\"a}tze minimize circuit depth to reduce noise accumulation, but may sacrifice expressibility compared to chemically-motivated ans{\"a}tze like unitary coupled cluster.

% Hardware-efficient ans{\"a}tze (HEA)~\cite{kandala2017hardware} are constructed from native gates of the target quantum processor, typically alternating layers of single-qubit rotations and two-qubit entangling gates:
% \begin{equation}
%     U(\boldsymbol{\theta}) = \prod_{l=1}^{L} \left[ U_{\text{ent}} \cdot \bigotimes_{i=1}^{n} R(\theta_{l,i}) \right],
%     \label{eq:hea}
% \end{equation}
% where $R(\theta)$ represents parameterized single-qubit rotations (e.g., $R_Y(\theta)$ or $R_Y(\theta)R_Z(\phi)$), $U_{\text{ent}}$ is a layer of entangling gates (e.g., CNOT or CZ in a nearest-neighbor or all-to-all pattern), and $L$ is the number of layers. 

%------------------------------------------------------------------------------
\subsection{Expressibility}
\label{subsec:expressibility}
%------------------------------------------------------------------------------

Expressibility quantifies how uniformly a parameterized quantum circuit can explore the Hilbert space~\cite{sim2019expressibility}. It is defined by comparing the distribution of state fidelities produced by a circuit to the distribution expected from Haar-random states. 
% For a parameterized circuit $U(\boldsymbol{\theta})$, consider pairs of states generated by independently sampled parameters:
% \begin{equation}
%     |\psi_{\boldsymbol{\theta}}\rangle = U(\boldsymbol{\theta})|0\rangle^{\otimes n}, \quad
%     |\psi_{\boldsymbol{\phi}}\rangle = U(\boldsymbol{\phi})|0\rangle^{\otimes n},
%     \label{eq:states}
% \end{equation}
% where $\boldsymbol{\theta}, \boldsymbol{\phi}$ are drawn uniformly from the parameter domain (typically $[0, 2\pi]^p$). The fidelity between these states is
% \begin{equation}
%     F = |\langle\psi_{\boldsymbol{\theta}}|\psi_{\boldsymbol{\phi}}\rangle|^2.
%     \label{eq:fidelity}
% \end{equation}
Sampling many parameter pairs yields an empirical distribution $\hat{P}_{\text{PQC}}$. For Haar-random states on an $N = 2^n$ dimensional Hilbert space, the fidelity distribution is known analytically, $P_{\text{Haar}}(F) = (N-1)(1-F)^{N-2}$.
% \begin{equation}
%     P_{\text{Haar}}(F) = (N-1)(1-F)^{N-2}.
%     \label{eq:haar}
% \end{equation}

Expressibility is computed as the Kullback-Leibler (KL) divergence between the circuit's fidelity distribution and the Haar distribution.
% \begin{equation}
%     \text{Expr} = D_{\text{KL}}\left(\hat{P}_{\text{PQC}}(F) \| P_{\text{Haar}}(F)\right) = \sum_{k} \hat{P}_{\text{PQC}}(F_k) \log \frac{\hat{P}_{\text{PQC}}(F_k)}{P_{\text{Haar}}(F_k)},
%     \label{eq:expressibility}
% \end{equation}
%where the sum is over histogram bins $k$ used to discretize the fidelity distributions. 
Lower expressibility values indicate closer agreement with the Haar distribution, meaning more uniform state-space coverage. A circuit achieving $\text{Expr} = 0$ would form an exact 2-design.

Expressibility provides a problem-independent characterization of a circuit's representational capacity. Circuits with low expressibility (high uniformity) can in principle represent a wider variety of target states. However, as discussed in Section~\ref{sec:related}, high expressibility correlates with barren plateaus and does not guarantee good VQE performance, particularly under noise.

%------------------------------------------------------------------------------
\subsection{Quantum Error Mitigation}
\label{subsec:qem}
%------------------------------------------------------------------------------

Quantum error mitigation (QEM) techniques aim to reduce the effect of noise on expectation value estimates without the overhead of full quantum error correction~\cite{cai2023quantum,temme2017error}. We focus on two widely-used techniques: zero-noise extrapolation (ZNE) and probabilistic error cancellation (PEC).

% %..............................................................................
% \subsubsection{Zero-Noise Extrapolation}
% \label{subsubsec:zne}
% %..............................................................................

Zero-noise extrapolation estimates the ideal (zero-noise) expectation value by artificially amplifying noise and extrapolating to the zero-noise limit~\cite{temme2017error,giurgica2020digital}.
The noise level is characterized by a scale factor $\lambda \geq 1$, where $\lambda = 1$ corresponds to the native hardware noise. Noise can be amplified to levels $\lambda > 1$ through several methods such as pulse stretching, unitary folding, or identity insertion.
% \begin{itemize}
%     \item Pulse stretching: Extending gate pulse durations by factor $\lambda$, increasing decoherence
%     \item Unitary folding: Replacing each gate $G$ with the sequence $G(G^\dagger G)^{(k)}$ where $\lambda = 2k + 1$
%     \item Identity insertion: Adding pairs of gates that multiply to identity, each introducing additional noise
% \end{itemize}
Unitary folding is most commonly used as it requires only gate-level access and applies to arbitrary circuits. 
%For a circuit $U = G_m \cdots G_2 G_1$, global folding replaces it with
% \begin{equation}
%     U \mapsto U(U^\dagger U)^k = G_m \cdots G_1 (G_1^\dagger \cdots G_m^\dagger G_m \cdots G_1)^k,
%     \label{eq:folding}
% \end{equation}
% effectively tripling the circuit depth for each fold ($k=1$ gives $\lambda=3$).

Given expectation values $\langle O \rangle_\lambda$ measured at multiple noise levels $\{\lambda_1, \lambda_2, \ldots, \lambda_m\}$, the zero-noise value $\langle O \rangle_{\lambda=0}$ is estimated by fitting and extrapolating. Common models include:
\begin{itemize}
    \item Linear: $\langle O \rangle_\lambda = a + b\lambda$, extrapolating to $\lambda = 0$
    \item Polynomial: $\langle O \rangle_\lambda = \sum_{j=0}^{d} a_j \lambda^j$ for degree $d$
    \item Exponential: $\langle O \rangle_\lambda = a + b e^{-c\lambda}$, appropriate for depolarizing noise
\end{itemize}

The Richardson extrapolation formula for $m$ noise levels uses $
    \langle O \rangle_{\text{ZNE}} = \sum_{i=1}^{m} \gamma_i \langle O \rangle_{\lambda_i},
$
where coefficients $\gamma_i$ are determined by the noise levels and extrapolation model. 
% For linear extrapolation with $\lambda_1 = 1$ and $\lambda_2 = 3$:
% \begin{equation}
%     \langle O \rangle_{\text{ZNE}} = \frac{3\langle O \rangle_1 - \langle O \rangle_3}{2}.
%     \label{eq:linear_zne}
% \end{equation}
% \paragraph{Overhead.}
ZNE requires $m$ circuit executions at different noise levels, each with the same number of shots. The sampling overhead is thus a constant factor $m$. However, noise amplification increases effective circuit depth, which may introduce additional errors for very deep circuits or degrade extrapolation accuracy when noise deviates from assumed models.

% %..............................................................................
% \subsubsection{Probabilistic Error Cancellation}
% \label{subsubsec:pec}
% %..............................................................................

Probabilistic error cancellation constructs an unbiased estimator of the ideal expectation value by representing ideal operations as quasi-probability distributions over noisy operations~\cite{temme2017error}.

% \paragraph{Quasi-Probability Representation.}
Consider an ideal gate $\mathcal{G}$ and its noisy implementation $\tilde{\mathcal{G}}$. If the noise channel $\mathcal{N}$ is known and invertible, we can write, $
    \mathcal{G} = \mathcal{N}^{-1} \circ \tilde{\mathcal{G}} = \sum_{i} \eta_i \mathcal{B}_i \circ \tilde{\mathcal{G}},
$
where $\{\mathcal{B}_i\}$ is a basis of implementable operations (typically Pauli operators) and $\eta_i$ are real coefficients that may be negative.
% \paragraph{Monte Carlo Estimation.}
Rather than implementing $\mathcal{N}^{-1}$ directly, PEC uses Monte Carlo sampling. For each circuit execution:
\begin{enumerate}
    \item For each gate, sample a correction operation $\mathcal{B}_i$ with probability $|\eta_i|/\gamma$, where $\gamma = \sum_i |\eta_i| \geq 1$
    \item Execute the circuit with sampled corrections inserted after each gate
    \item Weight the measurement outcome by $\text{sign}(\eta_i)$ for each sampled correction
\end{enumerate}

The mitigated estimator is $
    \langle O \rangle_{\text{PEC}} = \gamma^{N_g} \cdot \mathbb{E}\left[ s \cdot o \right],
$
where $N_g$ is the number of gates, $s = \prod_j \text{sign}(\eta_{i_j})$ is the product of signs for sampled corrections, and $o$ is the measurement outcome.
% \paragraph{Overhead.}
The variance of the PEC estimator scales as $\gamma^{2N_g}$, requiring $\gamma^{2N_g}$ times more samples than unmitigated estimation to achieve the same statistical precision. For typical noise rates ($\gamma \approx 1.01$--$1.1$ per gate), this overhead grows exponentially with circuit depth, making PEC practical only for relatively shallow circuits. The cost factor $\gamma$ depends on noise strength: lower noise yields $\gamma$ closer to 1 and lower overhead.

% \paragraph{Noise Characterization.}
PEC requires an accurate noise model to construct the quasi-probability representation. This is typically obtained through gate set tomography, randomized benchmarking, or cycle benchmarking. Sparse Pauli-Lindblad models~\cite{kim2023evidence} provide a scalable approach by assuming noise acts as a sparse combination of Pauli channels, reducing characterization overhead while maintaining accuracy.

% %..............................................................................
% \subsubsection{Comparison of ZNE and PEC}
% \label{subsubsec:comparison}
% %..............................................................................

Table~\ref{tab:qem_comparison} summarizes the key differences between ZNE and PEC relevant to our study.
ZNE is simpler to implement and has constant overhead, but relies on extrapolation model assumptions that may fail for complex noise. PEC provides theoretically unbiased estimates but requires detailed noise characterization and has exponential sampling overhead. In practice, both methods achieve significant error reduction for shallow-to-moderate depth circuits typical of near-term VQE applications.

\begin{table*}[ht]
\small
\centering
\caption{Comparison of zero-noise extrapolation (ZNE) and probabilistic error cancellation (PEC).}
\label{tab:qem_comparison}
\begin{tabular}{lll}
\hline
\textbf{Property} & \textbf{ZNE} & \textbf{PEC} \\
\hline
Bias & Biased (model-dependent) & Unbiased (infinite samples) \\
Sampling overhead & Constant ($\times m$) & Exponential ($\gamma^{2N_g}$) \\
Noise model required & No & Yes (detailed characterization) \\
Extrapolation model & Required & Not required \\
Circuit depth sensitivity & Moderate & High \\
Implementation complexity & Low & Moderate \\
\hline
\end{tabular}
\end{table*}

%==============================================================================
\section{Methods}
\label{sec:methods}
%==============================================================================

% %------------------------------------------------------------------------------
% \subsection{Molecular Systems}
% \label{subsec:molecules}
% %------------------------------------------------------------------------------

We study two molecules in depth, and four molecules in total, spanning 4 to 14 qubits:

\begin{enumerate}
    \item Hydrogen, \ce{H2}: 4 qubits, 12 ansatz circuits (circuits 1--12)
    \item Trihydrogen Cation, \ce{H3+}: 6 qubits, 6 ansatz circuits (circuits 19--24)
    \item (Scaling Only) Lithium Hydride, \ce{LiH}: 12 qubits, 4 ansatz circuits (circuits 13--14, 17--18)
    \item (Scaling Only) Beryllium Dihydride, \ce{BeH2}: 14 qubits, 2 ansatz circuits (circuits 25, 27)
\end{enumerate}

Molecular Hamiltonians were computed using PySCF \cite{sun2018pyscf} with the STO-3G basis set and mapped to qubit operators via Jordan-Wigner transformation using {\tt Qiskit Nature}. Exact ground state energies are $-1.1373$ Hartree for \ce{H2} (at $0.735$ \AA), $-1.5672$ Hartree for \ce{H3+} (equilateral triangle, $0.87$ \AA\ bond length), $-7.8823$ Hartree for \ce{LiH} (at $1.546$ \AA), and $-15.5952$ Hartree for \ce{BeH2} (linear, $1.326$ \AA\ bond length).

Full error mitigation experiments (IDEAL, COMBINED, ZNE, PEC) and noise model comparisons (DEPOL, AMPLITUDE\_DAMPING) were conducted for \ce{H2} and \ce{H3+}. For \ce{LiH} and \ce{BeH2}, only ideal (noiseless) experiments were feasible: noisy simulation at 12--14 qubits requires density matrix operations on $2^n \times 2^n$ matrices, making a single noisy VQE run computationally intractable. %(estimated $\sim$57 days for \ce{LiH}).
%These larger molecules serve as scaling reference points for expressibility analysis.

% Description of circuit families tested
We evaluate 24 hardware-efficient ansatz circuits across the four molecules with varying:
\begin{itemize}
    \item Single-qubit rotation types ($R_Y$, $R_Y R_Z$, $H R_X$)
    \item Two-qubit gate types (CNOT, CZ)
    \item Entanglement patterns (linear, circular, all-to-all)
    \item Circuit depth (1--2 layers)
\end{itemize}

For \ce{H2}, the 12 circuits span 4--6 two-qubit gates and depths 6--9. For \ce{H3+}, circuits have 5--15 two-qubit gates and depths 4--11, providing greater structural diversity for correlation analysis. \ce{LiH} circuits have 11--12 two-qubit gates and \ce{BeH2} circuits have 13--14, all with 24--28 parameters. \footnote{Circuit diagrams and code are provided at \url{https://github.com/evanccoleman/VQEsensitivity/releases/tag/v1-QUASAR-paper-release}}

% Noise model used
We simulate three noise models to test the robustness of our findings and to isolate the contribution of different noise channels:

\begin{enumerate}
    \item COMBINED (primary): Depolarizing error composed with thermal relaxation, matching Saib et al.'s~\cite{saib2021effect} methodology. Single-qubit gates ($R_Y$): depolarizing probability $p_1 = 0.005$, thermal relaxation $T_1 = 20\,\mu$s, $T_2 = 12\,\mu$s, gate time $70$\,ns. Two-qubit gates (CNOT): $p_2 = 0.035$, gate time $320$\,ns.
    \item DEPOL: Pure depolarizing error only (same $p_1$, $p_2$, no thermal relaxation). Under this model, PEC's depolarizing noise representation is \emph{exactly correct}, providing a controlled test of whether PEC failure stems from noise model mismatch or fundamental overhead.
    \item AMPLITUDE\_DAMPING: Pure amplitude damping ($T_1$ decay) with damping parameters derived from gate times and $T_1$. This non-unital channel tests the Singkanipa and Lidar~\cite{singkanipa2025beyond} prediction that non-unital noise does not necessarily induce barren plateaus.
\end{enumerate}

Following Saib et al., noise is applied only to $R_Y$ and CNOT gates to isolate the effect of parameterized and entangling operations. Simulations were performed using {\tt Qiskit Aer} \cite{javadi2024qiskit} with automatic simulation method selection.
VQE optimization was performed using the SPSA optimizer \cite{spall1998overview} with 200 maximum iterations. For \ce{H2} and \ce{H3+}, we performed 10 independent VQE runs per circuit--condition pair; for \ce{LiH} and \ce{BeH2}, we performed 3 runs due to longer per-run computation times. In all cases we report the lowest energy, which reduces sensitivity to optimizer stochasticity and isolates the effect of circuit architecture and noise. Run-to-run standard deviations are reported to characterize landscape ruggedness.

% Ground truth
Exact ground state energies were computed using NumPy's eigensolver for comparison. VQE performance is quantified by the energy error, $\Delta E = |E_{\text{VQE}} - E_{\text{exact}}|$.
All experiments used {\tt Qiskit 1.4.5, Qiskit Aer 0.17}, and {\tt Mitiq 0.41}. %\ce{H2} and \ce{H3+} experiments (180 total across all noise models and mitigation conditions) ran on a 32-core CPU cluster, completing in approximately 4 hours with 32-way parallelism. \ce{LiH} and \ce{BeH2} ideal experiments (6 total) ran on a GPU-accelerated workstation (NVIDIA RTX 5090) in approximately 5 hours.
% Computational details
Expressibility was computed for each ansatz by:
\begin{enumerate}
    \item Sampling $N = 5000$ pairs of random parameters
    \item Computing state fidelities (ideal: using Statevector; noisy: using density matrices)
    \item Constructing histograms with 75 bins over $[0,1]$
    \item Computing KL divergence from the Haar distribution
\end{enumerate}

Seeds were fixed for reproducibility. Both ideal expressibility (noiseless simulation) and effective expressibility (noisy simulation) were computed for each ansatz.

%------------------------------------------------------------------------------
\subsection{Error Mitigation Protocols}
\label{subsec:mitigation_protocols}
%------------------------------------------------------------------------------

We implemented ZNE using Mitiq~\cite{larose2022mitiq} with noise scaling set to global unitary folding with scale factors of $\{1.0, 1.5, 2.0\}$ and Richardson extrapolation. ZNE is applied as a post-processing step: after VQE optimization converges to optimal parameters $\boldsymbol{\theta}^*$ under noise, the optimal circuit is evaluated at multiple noise levels and extrapolated to the zero-noise limit.

PEC was implemented using Mitiq with depolarizing noise representations:
\begin{itemize}
    \item Noise representation: local depolarizing channel with uniform error rate $p_2 = 0.035$ applied to all gates
    \item Quasi-probability sampling: 200 samples per evaluation
\end{itemize}

\noindent\textbf{Note.} Our PEC implementation applies the two-qubit error rate uniformly to all gates, overestimating single-qubit gate noise by $7\times$. To isolate whether this causes PEC's poor performance, we run a controlled comparison under pure depolarizing noise where PEC's representation is exact (Section~\ref{subsec:noise_model_results}).

% %------------------------------------------------------------------------------
% \subsection{Correlation Analysis}
% \label{subsec:correlation}
% %------------------------------------------------------------------------------

\subsection{Analysis}
\label{sect:analysis}

We assess the relationship between expressibility and VQE performance using standard statistical techniques.
For comparison, we evaluate zero-cost circuit topology metrics as predictors of VQE performance:
\begin{itemize}
    \item Two-qubit gate count: Total number of CNOT/CZ gates
    \item Circuit depth: Number of sequential gate layers
    \item Expected infidelity: $1 - (1-p_1)^{n_{1q}}(1-p_2)^{n_{2q}}$, the analytical probability that at least one gate error occurs
    \item Noise-weighted gate count: $p_1 \cdot n_{1q} + p_2 \cdot n_{2q}$, a first-order approximation of total noise dose
\end{itemize}

%These metrics require no simulation and provide a baseline for assessing whether expensive expressibility computations add predictive value beyond what circuit structure alone reveals.

%==============================================================================
\section{Results}
\label{sec:results}
%==============================================================================

%------------------------------------------------------------------------------
\subsection{Ranking Instability Under Noise}
\label{subsec:ranking_results}
%------------------------------------------------------------------------------

We first reproduce and extend Saib et al.'s central finding: circuit rankings change dramatically under noise. \Cref{tab:metric_comparison} reports the full correlation matrix; here we highlight the key results.

Under ideal conditions, all 12 \ce{H2} circuits achieve similar accuracy ($|\Delta E| = 0.018$--$0.020$ Hartree), with the exception of circuit 10 ($|\Delta E| = 0.612$), which consistently fails to converge. Under noise, performance degrades unevenly: circuit 2 emerges as the best performer ($|\Delta E| = 0.039$) despite ranking 9th ideally, while circuit 12 drops from best performing to 10th. The Spearman rank correlation between ideal and noisy rankings is $\rho = -0.098$ ($p = 0.76$), confirming Saib et al.'s finding of essentially zero rank preservation.

For \ce{H3+}, the pattern intensifies. Circuit 21 is the best ideal performer ($|\Delta E| = 0.018$) but becomes the worst under noise ($|\Delta E| = 0.398$). This is driven by its 15 two-qubit gates accumulating substantially more noise than the 5--6 gates in circuits 20 and 23. The ideal-to-noisy rank correlation is $\rho = -0.257$, again showing no predictive relationship.

\begin{figure}[t]
    \centering
    \includegraphics[width=\columnwidth]{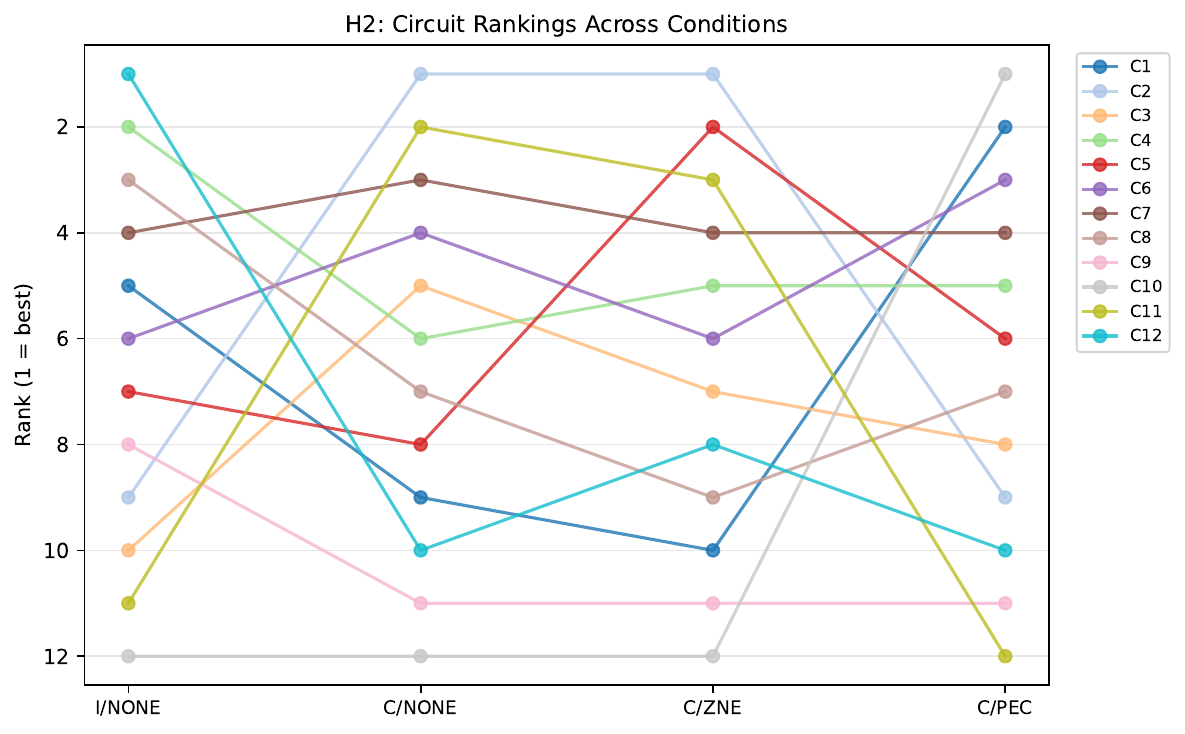}
    \caption{Circuit rankings across conditions for \ce{H2} (1 = best). Rankings scramble from Ideal to Noisy ($\rho = -0.10$), are largely preserved under ZNE ($\rho = +0.80$), and are actively reordered by PEC ($\rho = -0.22$).}
    \label{fig:ranking_bump}
\end{figure}

\Cref{fig:ranking_bump} visualizes this ranking instability for \ce{H2}. The Ideal to Noisy transition shows pervasive line crossings, while Noisy to ZNE lines remain largely parallel ($\rho = +0.80$). The Noisy to PEC transition reorders circuits unpredictably, with circuit 10 rising from last place to first; the sole case where PEC helps. This ranking instability is largely mirrored for \ce{H3+} which can be seen in \Cref{tab:results_h3p}.

%------------------------------------------------------------------------------
\subsection{Error Mitigation}
\label{subsec:mitigation_results}
%------------------------------------------------------------------------------

Contrary to expectations, error mitigation predominantly degrades rather than improves VQE performance.
For \ce{H2}, ZNE reduces error in only 4 of 12 circuits, with the mean error increasing from $0.140$ to $0.181$ Hartree. For \ce{H3+}, ZNE helps 4 of 6 circuits, with the mean error decreasing from $0.174$ to $0.132$ Hartree. The most dramatic improvement is circuit 21 (\ce{H3+}), where ZNE reduces error by 86\% ($0.398 \to 0.056$). 
PEC increases error in 11 of 12 \ce{H2} circuits and all 6 \ce{H3+} circuits. The mean \ce{H2} error rises from $0.140$ to $0.367$ Hartree; a $2.6\times$ increase. For \ce{H3+}, the mean rises from $0.174$ to $1.541$ Hartree, with individual degradations exceeding $4750\%$ (circuit 22). 
The only exception is \ce{H2} circuit 10, where PEC reduces error by 77\% ($0.609 \to 0.138$). This circuit fails to converge even ideally ($|\Delta E| = 0.612$), likely using PEC's quasi-probability sampling to escape local minima that standard VQE cannot.

\begin{figure}[t]
    \centering
    \includegraphics[width=\columnwidth]{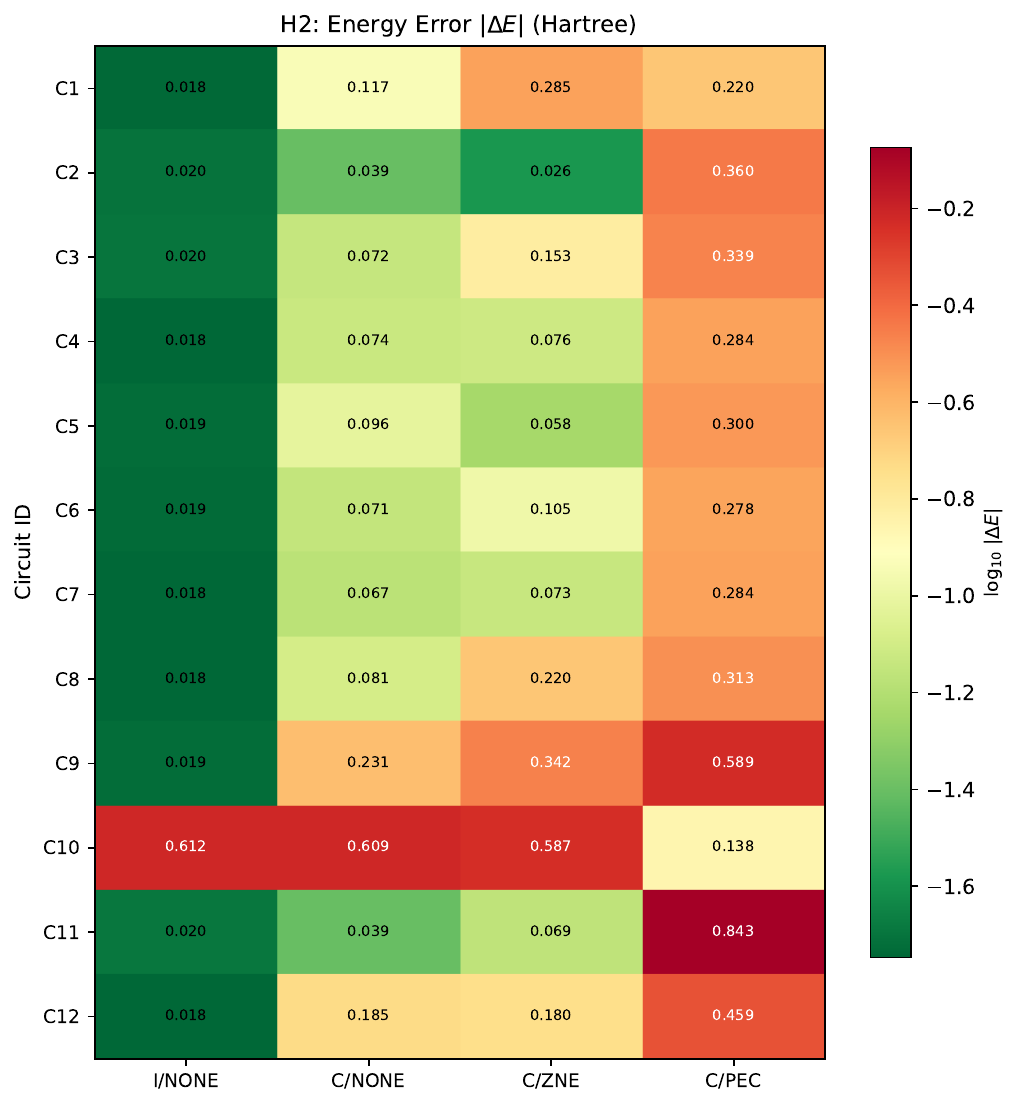}
    \caption{Energy error $|\Delta E|$ (Hartree) for all 12 \ce{H2} circuits across four conditions. Color scale is logarithmic; cell values show exact errors. PEC increases error for 11 of 12 circuits, often by $>$200\%.}
    \label{fig:heatmap}
\end{figure}

\Cref{fig:heatmap} summarizes these results. The Ideal column is uniformly green (low error), while the PEC column is almost entirely red.

%------------------------------------------------------------------------------
\subsection{Expressibility as a Predictor}
\label{subsec:expressibility_results}
%------------------------------------------------------------------------------

Standard expressibility (computed from ideal statevector simulations) shows moderate-to-strong correlation with VQE error under noisy and ZNE conditions for \ce{H2}: $r = +0.74$ ($p = 0.006$) for noisy and $r = +0.77$ ($p = 0.003$) for ZNE. This is substantially stronger than Saib et al.'s reported $r = 0.012$, likely because our best-of-10 methodology reduces optimizer stochasticity. However, standard expressibility shows no significant correlation for \ce{H3+} under any condition ($|r| < 0.38$, all $p > 0.4$), suggesting its predictive power does not generalize across molecules.

Noisy expressibility (computed from density matrix simulations at $O\left(4^n\right)$ cost) provides the strongest single predictor of unmitigated performance for \ce{H3+} ($r = +0.91$, $p = 0.01$) within this small set of circuits, but this result must be interpreted cautiously due to the small sample size. As discussed in Section~\ref{sec:discussion}, the pure-state Haar reference distribution used in the KL divergence is theoretically inappropriate for mixed states, and the resulting metric conflates expressibility loss with purity loss.

Neither expressibility metric predicts PEC performance for either molecule ($|r| < 0.31$, all $p > 0.26$), consistent with PEC degradation being driven by circuit complexity rather than state-space coverage.

%------------------------------------------------------------------------------
\subsection{Topology Metrics vs.\ Expressibility}
\label{subsec:targeted_efficiency}
%------------------------------------------------------------------------------

Table~\ref{tab:metric_comparison} compares ten candidate predictors across all conditions. Several zero-cost topology metrics are perfectly collinear for our circuit families and produce identical correlations.

\begin{figure}[t]
    \centering
    \includegraphics[width=\columnwidth]{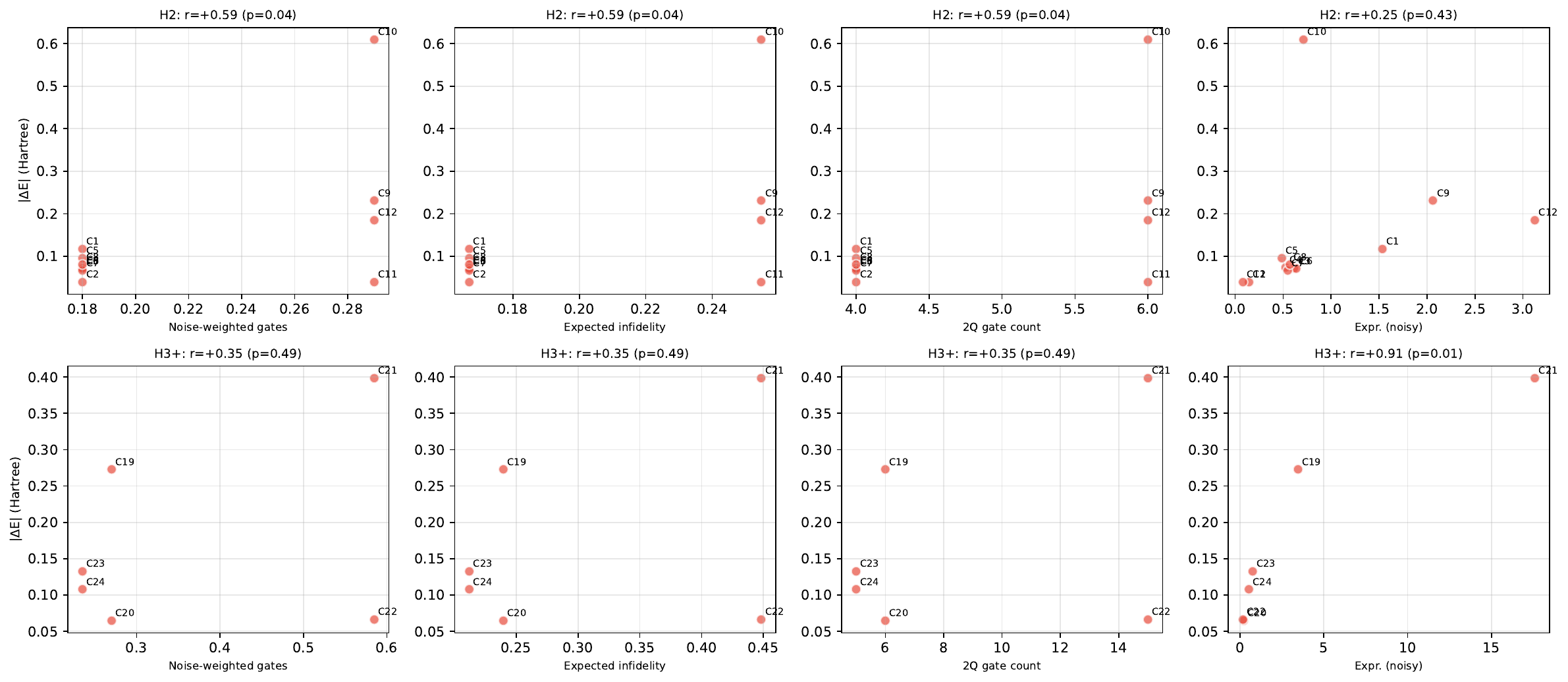}
    \caption{Scatter plots of four candidate metrics against noisy VQE error ($|\Delta E|$, COMBINED/NONE) for \ce{H2} (top) and \ce{H3+} (bottom). Noise-weighted gates, expected infidelity, and 2Q gate count all produce identical correlations within each molecule due to collinearity. Noisy expressibility is the strongest predictor for \ce{H3+} ($r = +0.91$) but adds little for \ce{H2} ($r = +0.25$).}
    \label{fig:metric_scatter}
\end{figure}

For PEC degradation, topology metrics are near-perfect predictors: two-qubit gate count achieves $r = +0.96$ for \ce{H3+}, reflecting the exponential scaling of PEC's quasi-probability overhead with gate count. For noisy baseline performance, the picture is more nuanced. The free metrics achieve only moderate correlations ($r \approx 0.35$--$0.59$) because \ce{H2} circuits 1--8 share identical gate counts, forcing all variation within this group to be explained by structural differences captured only by expressibility. Note that the \ce{H3+} results should be interpreted cautiously given the small sample size (n = 6).

%The key practical finding: no single metric dominates across all conditions. Standard expressibility best predicts noisy and ZNE performance for \ce{H2}, noisy expressibility best predicts noisy performance for \ce{H3+}, and gate count best predicts PEC degradation for both molecules. This condition-dependence has direct implications for ansatz selection, which we discuss in Section~\ref{sec:discussion}.

%------------------------------------------------------------------------------
\subsection{Noise Model Sensitivity}
\label{subsec:noise_model_results}
%------------------------------------------------------------------------------

To test whether our findings are artifacts of a particular noise model, we repeat all \ce{H2} and \ce{H3+} experiments under two additional noise channels: pure depolarizing (DEPOL) and amplitude damping (AMP\_DAMP). This also lets us test whether PEC's poor performance comes from representation mismatch or not.
%This also enables a critical ablation: under DEPOL, PEC's depolarizing noise representation is exactly correct, isolating whether PEC's poor performance stems from representation mismatch or fundamental overhead.

Under pure depolarizing noise, PEC helps only 1 of 12 \ce{H2} circuits and 0 of 6 \ce{H3+} circuits which is statistically identical to its performance under COMBINED noise. The mean \ce{H2} error increases from $0.121$ to $0.440$ Hartree ($+263\%$), worse than under COMBINED ($+162\%$). For \ce{H3+}, PEC under DEPOL degrades error by $1{,}137\%$ on average. Under amplitude damping, where PEC's representation is maximally mismatched, PEC is even more destructive: $+536\%$ mean degradation for \ce{H2} and $+2{,}621\%$ for \ce{H3+}. These results confirm that PEC's failure is driven primarily by exponential sampling overhead, not by noise model mismatch in our implementation. At $p_2 = 0.035$ the per-gate cost factor is $\gamma \approx 1.15$, giving variance overhead of $\gamma^{2N_g} \approx 30\times$ for 12-gate circuits and $\approx 66\times$ for 15-gate circuits, consistent with observed degradation regardless of whether the representation is exact.

ZNE performance is noise-model-dependent. Under DEPOL, ZNE helps 6 of 12 \ce{H2} circuits (vs. 4 under COMBINED), with mean error increasing only $10\%$ (vs.\ $29\%$ under COMBINED). Under amplitude damping, where baseline errors are already lower ($0.079$ vs.\ $0.140$ Hartree), ZNE helps 7 of 12 \ce{H2} circuits and reduces mean error by $2\%$. For \ce{H3+}, ZNE reduces mean error under all three noise models ($+24\%$ COMBINED, $+27\%$ DEPOL, $+37\%$ AMP\_DAMP). The improvement under amplitude damping is consistent with Singkanipa and Lidar's~\cite{singkanipa2025beyond} finding that non-unital noise is less detrimental to variational circuit training.

Circuit vulnerability ordering is remarkably stable across noise models. For \ce{H3+}, the Spearman rank correlation between noisy baselines is $\rho = 1.0$ for all three pairwise comparisons (COMBINED vs.\ DEPOL, COMBINED vs.\ AMP\_DAMP, DEPOL vs.\ AMP\_DAMP). For \ce{H2}, correlations range from $\rho = +0.78$ to $+0.87$ (all $p < 0.005$). This means that while absolute error magnitudes vary across noise models, the relative ranking of circuits is preserved---a circuit that is vulnerable to one noise type is vulnerable to all.
Standard expressibility maintains significant positive correlation with noisy baseline error across all three noise models for \ce{H2}: $r = +0.74$ (COMBINED), $r = +0.68$ (DEPOL), $r = +0.66$ (AMP\_DAMP), all $p < 0.03$. PEC error remains uncorrelated with expressibility regardless of noise model ($|r| < 0.25$). This consistency strengthens the case that standard expressibility captures a genuine structural property relevant to noisy VQE performance, rather than a noise-model-specific artifact.

\begin{table*}[htbp]
\small
\centering
\caption{Pearson correlation ($r$) between circuit metrics and VQE energy error $|\Delta E|$.
Metrics above the line are computable from circuit topology alone (zero simulation cost).
Bold indicates $|r| > 0.7$; $^*$ indicates $p < 0.05$.}
\label{tab:metric_comparison}
\begin{tabular}{llcccccccc}
\toprule
 & & \multicolumn{4}{c}{\ce{H2} (12 circuits)} & \multicolumn{4}{c}{\ce{H3+} (6 circuits)} \\
\cmidrule(lr){3-6} \cmidrule(lr){7-10}
Metric & Cost & Ideal & Noisy & ZNE & PEC & Ideal & Noisy & ZNE & PEC \\
\midrule
2Q gate count$^\dagger$ & Free & $+0.43$ & $+0.59^*$ & $+0.52$ & $+0.55$ & $-0.65$ & $+0.35$ & $-0.40$ & $\mathbf{+0.95}^*$ \\
Circuit depth & Free & $+0.43$ & $+0.59^*$ & $+0.52$ & $+0.55$ & $-0.34$ & $+0.37$ & $-0.38$ & $\mathbf{+0.82}^*$ \\
Params/qubit & Free & $-0.13$ & $-0.08$ & $-0.17$ & $\mathbf{+0.70}^*$ & -- & -- & -- & -- \\
\midrule
Expressibility (std) & $O\left(2^n\right)$ & $+0.63^*$ & $\mathbf{+0.74}^*$ & $\mathbf{+0.77}^*$ & $-0.25$ & $-0.08$ & $-0.38$ & $-0.22$ & $-0.55$ \\
Expressibility (noisy) & $O\left(4^n\right)$ & $-0.08$ & $+0.25$ & $+0.34$ & $+0.09$ & $-0.58$ & $\mathbf{+0.91}^*$ & $-0.34$ & $+0.31$ \\
\bottomrule
\multicolumn{10}{l}{\footnotesize $^\dagger$Total gates, noise-weighted gates, expected infidelity, $-\log$(survival), and 2Q gates/qubit are perfectly collinear} \\
\multicolumn{10}{l}{\footnotesize \phantom{$^\dagger$}with 2Q gate count for our circuit families and produce identical correlations.} \\
\end{tabular}
\end{table*}
%------------------------------------------------------------------------------
\subsection{Molecular Scaling}
\label{subsec:scaling}
%------------------------------------------------------------------------------

To investigate whether expressibility-performance relationships persist at larger scales, we perform a preliminary scaling study, where we run ideal (noiseless) VQE for \ce{LiH} (12 qubits, 4 circuits) and \ce{BeH2} (14 qubits, 2 circuits) with 3 independent runs per circuit. %Noisy simulation is computationally intractable at these scales: density matrix operations on $2^{12} \times 2^{12}$ matrices make a single noisy \ce{LiH} VQE run estimated at $\sim$57 days, compared to $\sim$20 minutes for ideal statevector simulation. This computational wall itself is a finding---error mitigation techniques that require noisy simulation cannot be applied at 12+ qubits in reasonable timeframes, reinforcing the need for cheap predictors of circuit quality.
Run-to-run standard deviations increase dramatically across molecules: $0.04$ Hartree (\ce{H2}, 4 qubits), $0.06$ (\ce{H3+}, 6 qubits), $0.15$ (\ce{LiH}, 12 qubits), and $0.32$ (\ce{BeH2}, 14 qubits). The best-of-3 ideal energy errors for \ce{LiH} range from $0.259$ to $0.492$ Hartree, far above chemical accuracy ($1.6 \times 10^{-3}$ Hartree), and \ce{BeH2} errors exceed $0.7$ Hartree. This confirms that 200 SPSA iterations with 3 runs is insufficient for 12+ qubit hardware-efficient ans{\"a}tze, consistent with the barren plateau predictions of Holmes et al.~\cite{holmes2022connecting}.

Standard expressibility values decrease sharply with qubit count. \ce{H2} circuits span $0.02$--$0.71$; \ce{H3+} circuits cluster at $0.20$--$0.28$; \ce{LiH} circuits fall below $0.05$; and both \ce{BeH2} circuits are near zero ($< 0.015$). As circuits approach 2-designs (expressibility $\to 0$), they become maximally expressive but correspondingly untrainable. This expressibility collapse means the metric loses discriminative power at larger scales since all circuits look equally expressive, yet their VQE performance still varies by a factor of two.
While standard expressibility correlates with ideal VQE error within \ce{H2} ($r = +0.63$, $p = 0.03$), it shows no predictive power across the full set of 24 circuits spanning all four molecules ($r = -0.18$, $p = 0.40$). A circuit with expressibility $0.02$ can have $|\Delta E| = 0.018$ (\ce{H2} circuit 12) or $|\Delta E| = 0.49$ (\ce{LiH} circuit 13). Qubit count and parameter space dimensionality overwhelm whatever structural information expressibility captures, confirming that expressibility is a within-molecule, within-scale metric rather than a universal predictor of VQE performance.

%==============================================================================
\section{Discussion}
\label{sec:discussion}
%==============================================================================

While noisy expressibility demonstrates a moderate correlation with unmitigated VQE performance, interpreting this correlation requires careful physical distinction. Under significant noise, the accessible Hilbert space collapses inward toward the maximally mixed state. Consequently, the Kullback-Leibler (KL) divergence from the pure-state Haar distribution inflates dramatically not because the ans{a}tz lacks the architectural capacity to explore the state space, but because the noise channel has physically restricted the purity of the sampled states. In this regime, effective expressibility ceases to be a pure measure of representational capacity and instead morphs into a metric of noise accumulation. Because this purity loss scales reliably with circuit depth and entangling gate count, the noisy KL divergence naturally tracks with VQE error. However, computing this metric requires computationally expensive density matrix simulations across thousands of parameter samples. Given that straightforward hardware efficiency metrics such as two-qubit gate count provide a direct, zero-cost proxy for the exact same noise susceptibility, the utility of effective expressibility for practical ansatz selection in the NISQ era is largely redundant.

PEC's poor performance is attributable to exponential sampling overhead, not noise model mismatch. Our ablation study (\Cref{subsec:noise_model_results}) shows that PEC under pure depolarizing noise, where its quasi-probability representation is exact, performs no better than under COMBINED noise, and actually performs worse in absolute terms ($+263\%$ vs.\ $+162\%$ mean degradation for \ce{H2}). This rules out our uniform noise representation as the primary cause. The fundamental issue is PEC's variance scales as $\gamma^{2N_g}$, and for circuits with $10$--$21$ gates at noise rates $p_2 = 0.035$, this overhead produces variance that overwhelms the bias correction. The strong correlation between gate count and PEC degradation ($r = +0.96$ for \ce{H3+}, consistent across all noise models) reflects this exponential scaling.

ZNE's mixed results reflect its dependence on the noise-energy relationship being smooth and monotonic. When noise scaling by unitary folding produces a well-behaved curve, Richardson extrapolation accurately recovers the zero-noise limit. For circuits where folding changes the effective noise character (e.g., by introducing coherent error accumulation), extrapolation can overshoot. This noise dependence is confirmed by the noise model comparison: ZNE helps 7 of 12 \ce{H2} circuits under amplitude damping versus 4 under COMBINED, consistent with Singkanipa and Lidar's prediction that non-unital noise is less detrimental to variational circuits. Separately, the molecule dependence (ZNE helps 4/6 \ce{H3+} circuits but only 4/12 \ce{H2} circuits) may reflect \ce{H3+}'s deeper circuits having more room for noise amplification to produce a clean extrapolation.

The strong correlation between noisy expressibility and unmitigated performance for \ce{H3+} ($r = +0.91$) is empirically striking but theoretically fragile. The Haar fidelity distribution is derived for pure states; under noise, circuits produce mixed states whose fidelities are compressed toward lower values due to purity loss, not reduced state-space coverage. The resulting KL divergence conflates two physically distinct phenomena: (1) the circuit's inability to uniformly explore the Hilbert space (true expressibility loss) and (2) noise-induced decoherence reducing state purity. Because both scale with circuit depth and gate count, the noisy KL divergence naturally tracks VQE error, but it does so for the same reason that gate count does. The practical implication is that noisy expressibility adds predictive value primarily when topology metrics cannot distinguish circuits.

%Our results suggest a tiered selection strategy. First, eliminate circuits with high gate counts (especially two-qubit gates), as these will perform poorly under noise regardless of mitigation. Among circuits with similar gate counts, standard expressibility distinguishes which are more likely to succeed under noisy or ZNE conditions. Noisy expressibility adds further discrimination for small molecules where density matrix simulation is tractable, but offers diminishing returns as system size increases. For PEC specifically, gate count alone is nearly sufficient to predict performance.

%==============================================================================
\section{Conclusion \& Future Directions}
\label{sec:conclusion}
%==============================================================================

We have investigated the relationship between expressibility and VQE performance across \ce{H2} and \ce{H3+} under ideal, noisy, ZNE, and PEC conditions, with three noise models. Our findings challenge the hypothesis that error mitigation restores expressibility as a universal ansatz selection metric. Instead, we find that (1)~PEC overwhelmingly degrades performance regardless of noise model, including under pure depolarizing noise where its representation is exact, confirming that the failure is driven by fundamental sampling overhead; (2)~ZNE provides circuit and noise dependent improvement, performing best under amplitude damping; (3)~circuit rankings under noise are remarkably stable across noise models (\ce{H3+}: $\rho = 1.0$); (4)~standard expressibility significantly predicts noisy and ZNE performance for \ce{H2} ($r = 0.66$--$0.77$) across all noise models but not for larger molecules; (5)~in a preliminary scaling study on \ce{LiH} and \ce{BeH2}, expressibility values collapse toward zero at 12--14 qubits as circuits approach 2-designs, losing discriminative power; and (6)~noisy simulation is computationally intractable beyond $\sim$10 qubits, making error mitigation and noisy expressibility infeasible at practical scales.

For practitioners, these results suggest a practical tiered strategy: first filter by gate count to eliminate noise-vulnerable circuits, then use standard expressibility (which is cheap to compute) to rank the remaining candidates. Noisy expressibility adds value for small systems but becomes computationally intractable beyond $\sim$10 qubits.

Several directions merit further investigation:
\begin{enumerate}
    \item Per-gate PEC representations: Implementing gate-type-specific noise representations to determine how much of PEC's degradation stems from our uniform approximation versus fundamental overhead scaling
    \item Larger molecules: Extending the noisy baseline analysis to \ce{LiH} (12 qubits) and \ce{BeH2} (14 qubits), where mitigation is computationally intractable but expressibility-performance correlations can still be assessed using novel techniques
    \item Sampling-based noisy expressibility: Developing $O\left(2^n\right)$ approximations to noisy expressibility using measurement-based fidelity estimation rather than full density matrix simulation
    \item Real hardware validation: Confirming that simulation-based findings transfer to actual quantum processors with device-specific noise profiles
\end{enumerate}

%==============================================================================
% Acknowledgments
%==============================================================================
\begin{acks}
This work leveraged the ACES Cluster at Texas A\&M University under allocation CIS250436 from the Advanced Cyberinfrastructure Coordination Ecosystem: Services \& Support (ACCESS) program, which is supported by U.S. National Science Foundation grants \#2138259, \#2138286, \#2138307, \#2137603, and \#2138296.
\end{acks}

%==============================================================================
% Bibliography
%==============================================================================
\bibliographystyle{ACM-Reference-Format}
\bibliography{references}

%==============================================================================
% Appendices
%==============================================================================
\appendix

\section{Circuit Diagrams}
\label{app:circuits}

\begin{figure*}[p]
    \centering
    \includegraphics[width=\textwidth]{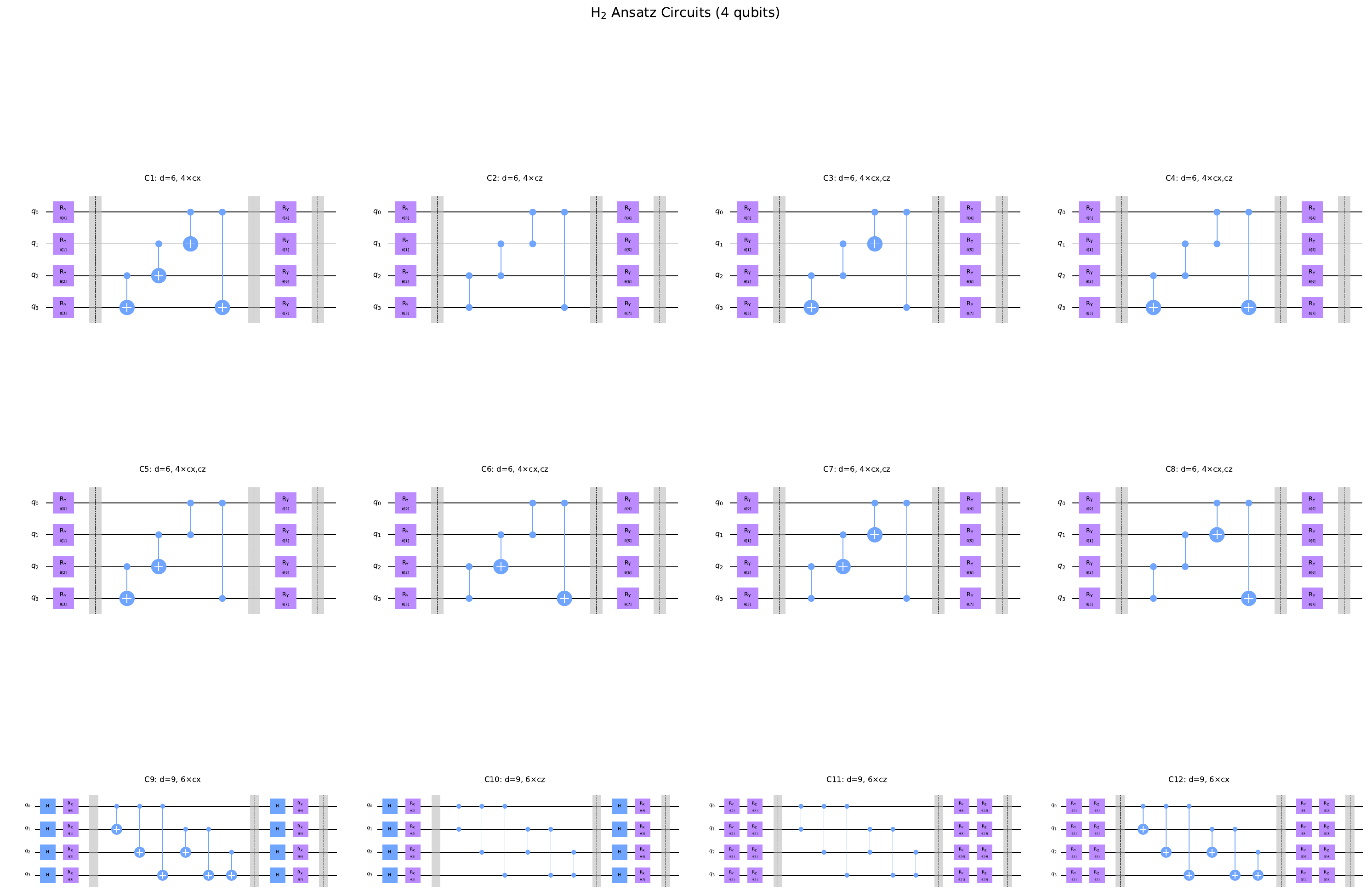}
    \caption{All 12 \ce{H2} ansatz circuits (4 qubits). Circuits 1--8 share the same gate count (8$\times$1Q, 4$\times$2Q) but differ in gate types and entanglement patterns. Circuits 9--12 use two layers (16$\times$1Q, 6$\times$2Q).}
    \label{fig:circuits_h2}
\end{figure*}

\begin{figure*}[p]
    \centering
    \includegraphics[width=\textwidth]{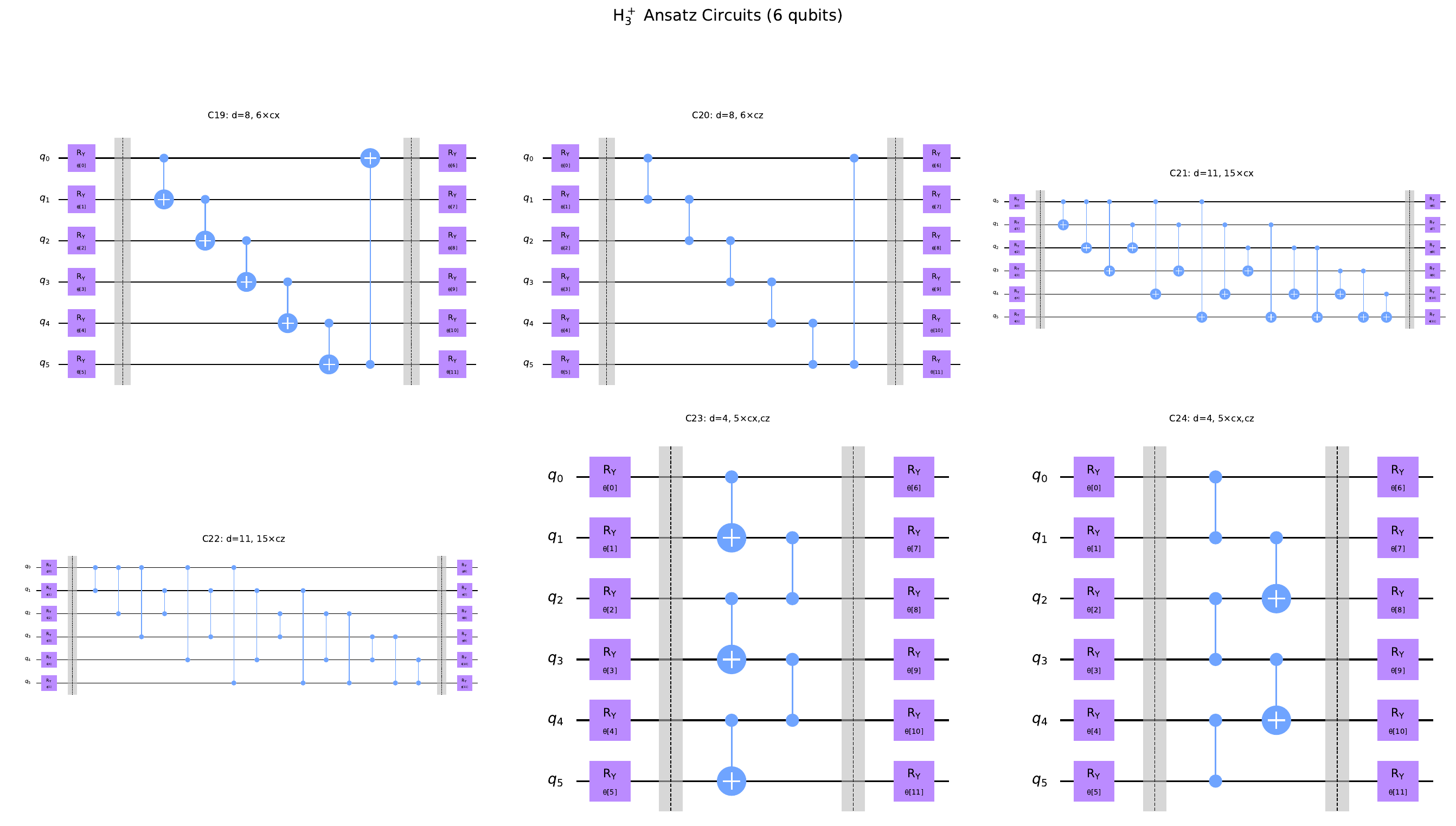}
    \caption{All 6 \ce{H3+} ansatz circuits (6 qubits). Circuits 21--22 have 15 two-qubit gates (all-to-all entanglement), while circuits 23--24 have only 5 (alternating pattern).}
    \label{fig:circuits_h3}
\end{figure*}

\begin{figure*}[htbp]
    \centering
    \begin{minipage}[t]{0.48\columnwidth}
        \centering
        \includegraphics[width=\linewidth]{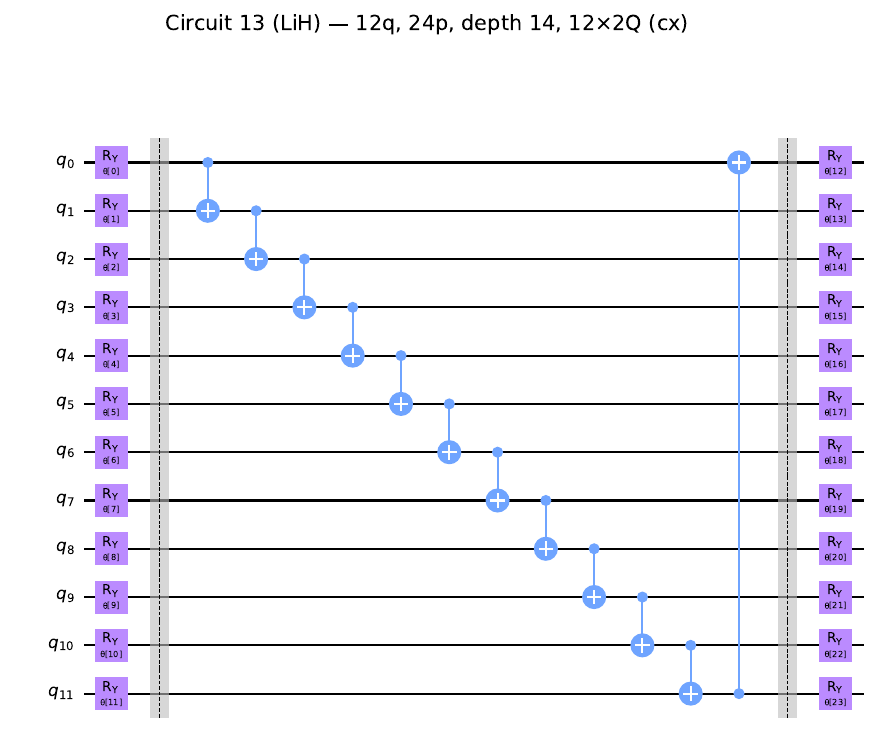}
    \end{minipage}\hfill
    \begin{minipage}[t]{0.48\columnwidth}
        \centering
        \includegraphics[width=\linewidth]{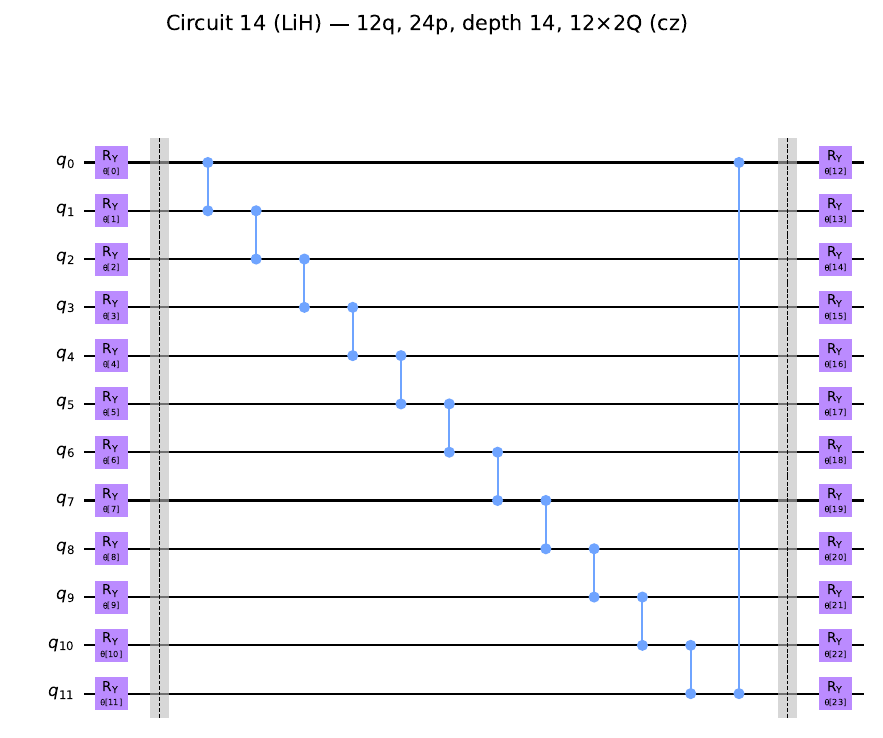}
    \end{minipage}
    
    \vspace{0.5em}
    
    \begin{minipage}[t]{0.48\columnwidth}
        \centering
        \includegraphics[width=\linewidth]{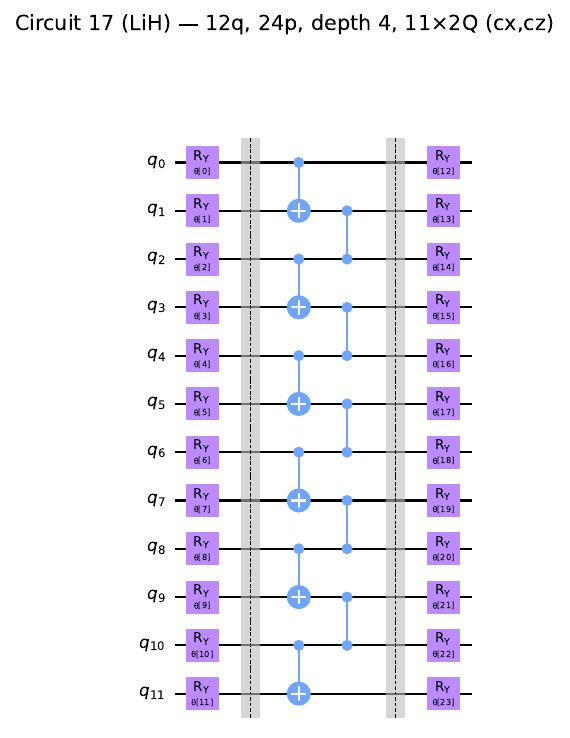}
    \end{minipage}\hfill
    \begin{minipage}[t]{0.48\columnwidth}
        \centering
        \includegraphics[width=\linewidth]{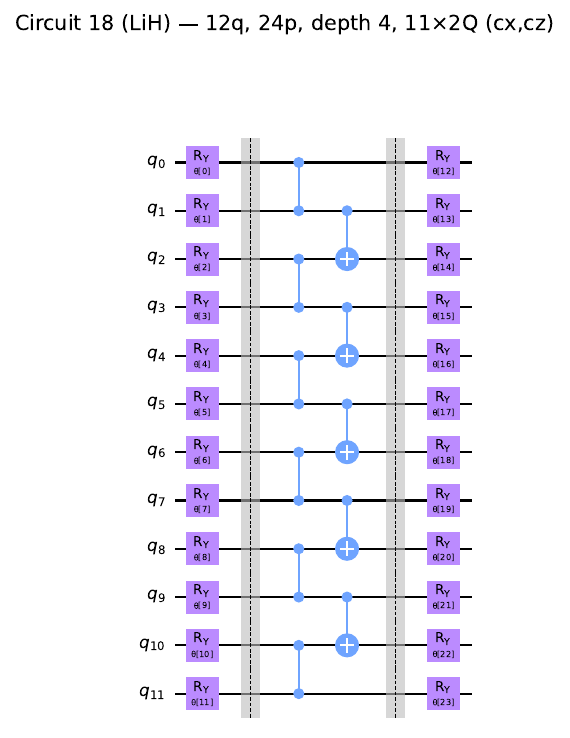}
    \end{minipage}
    \caption{\ce{LiH} ansatz circuits (12 qubits). Circuits 13--14 use ring entanglement (depth 14), circuits 17--18 use alternating entanglement (depth 4).}
    \label{fig:circuits_LiH}
\end{figure*}

\begin{figure*}[htbp]
    \centering
    \begin{minipage}[t]{0.48\columnwidth}
        \centering
        \includegraphics[width=\linewidth]{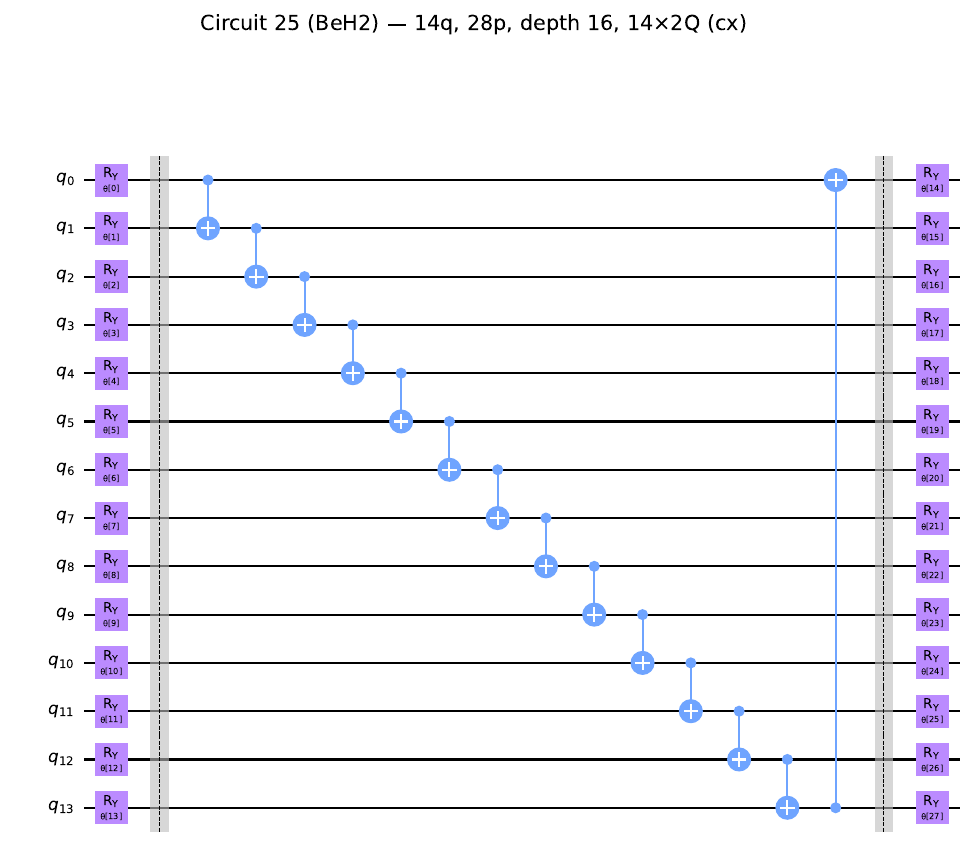}
    \end{minipage}\hfill
    \begin{minipage}[t]{0.48\columnwidth}
        \centering
        \includegraphics[width=\linewidth]{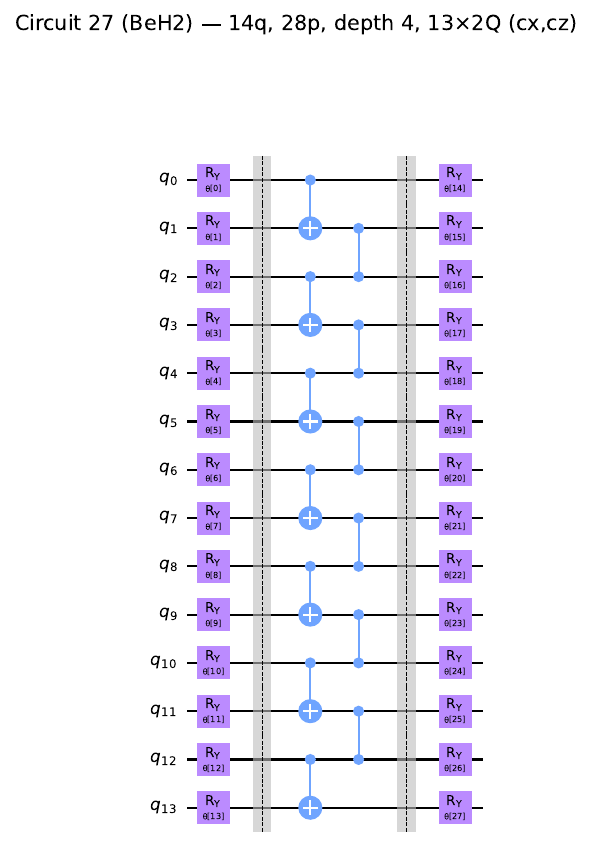}
    \end{minipage}
    \caption{\ce{BeH2} ansatz circuits (14 qubits). Circuit 25 uses ring entanglement (depth 16, 14$\times$CX), circuit 27 uses alternating entanglement (depth 4, 13$\times$CX).}
    \label{fig:circuits_beh2}
\end{figure*}

\Cref{fig:circuits_h2,fig:circuits_h3} show all ansatz circuits evaluated for \ce{H2} and \ce{H3+}, respectively. Individual \ce{LiH} and \ce{BeH2} circuit diagrams are shown in \Cref{fig:circuits_LiH,fig:circuits_beh2}.

\section{Detailed Numerical Results}

\begin{table}[htbp]
\centering
\caption{VQE Energy Error $|\Delta E|$ (Hartree) for \ce{H2}}
\label{tab:results_h2}
\begin{tabular}{lcccc}
\toprule
Circuit & I/NONE & C/NONE & C/ZNE & C/PEC \\
\midrule
C2 & 0.0197 & 0.0393 & 0.0264 & 0.3605 \\
C11 & 0.0202 & 0.0393 & 0.0686 & 0.8433 \\
C7 & 0.0180 & 0.0667 & 0.0728 & 0.2841 \\
C6 & 0.0185 & 0.0713 & 0.1052 & 0.2785 \\
C3 & 0.0201 & 0.0716 & 0.1528 & 0.3386 \\
C4 & 0.0180 & 0.0738 & 0.0761 & 0.2844 \\
C8 & 0.0180 & 0.0807 & 0.2204 & 0.3128 \\
C5 & 0.0189 & 0.0957 & 0.0576 & 0.2999 \\
C1 & 0.0183 & 0.1172 & 0.2850 & 0.2202 \\
C12 & 0.0179 & 0.1848 & 0.1801 & 0.4586 \\
C9 & 0.0189 & 0.2312 & 0.3421 & 0.5895 \\
C10 & 0.6117 & 0.6092 & 0.5868 & 0.1376 \\
\bottomrule
\end{tabular}
\end{table}

\begin{table}[htbp]
\centering
\caption{VQE Energy Error $|\Delta E|$ (Hartree) for \ce{H3+}}
\label{tab:results_h3p}
\begin{tabular}{lcccc}
\toprule
Circuit & I/NONE & C/NONE & C/ZNE & C/PEC \\
\midrule
C20 & 0.0352 & 0.0648 & 0.0517 & 1.1412 \\
C22 & 0.0257 & 0.0663 & 0.1240 & 3.2160 \\
C24 & 0.0363 & 0.1080 & 0.2316 & 0.9089 \\
C23 & 0.0225 & 0.1326 & 0.0926 & 0.8117 \\
C19 & 0.0389 & 0.2730 & 0.2359 & 0.8371 \\
C21 & 0.0178 & 0.3984 & 0.0556 & 2.3323 \\
\bottomrule
\end{tabular}
\end{table}

\begin{table}[htbp]
\centering
\caption{VQE Energy Error $|\Delta E|$ (Hartree) for \ce{LiH}}
\label{tab:results_LiH}
\begin{tabular}{lc}
\toprule
Circuit & I/NONE \\
\midrule
C14 & 0.2594 \\
C18 & 0.2595 \\
C17 & 0.3382 \\
C13 & 0.7006 \\
\bottomrule
\end{tabular}
\end{table}

\begin{table}[htbp]
\centering
\caption{VQE Energy Error $|\Delta E|$ (Hartree) for \ce{BeH2}}
\label{tab:results_beh2}
\begin{tabular}{lc}
\toprule
Circuit & I/NONE \\
\midrule
C25 & 0.7284 \\
C27 & 0.7704 \\
\bottomrule
\end{tabular}
\end{table}

\end{document}